\documentclass{ws-procs961x669}            
\begin{document}
\title{Indirect search for dark matter with neutrino telescopes}

\author{J. D. Zornoza}

\address{IFIC (Universidad de Valencia - CSIC)\\ c/ Catedr\'{a}tico Jos\'{e} Belrtr\'{a}n, 2 \\ Paterna 46980, Valencia (Spain)\\$^*$E-mail: zornoza@ific.uv.es}

\begin{abstract}
The quest to understand the nature dark matter is one of the most relevant ones in Particle Physics nowadays, since it constitutes most of the matter of the Universe and it is still unknown what it is made of. In order to answer to this question, a multi-front attack is needed because our knowledge of its properties is very incomplete. Among the different experimental strategies, neutrino telescopes are very relevant tools. There are several promising sources to look at: the Sun, the Galactic Center, the Earth, dwarf galaxies, galaxy clusters... As an example of the power of neutrino telescopes, we can mention the analysis of the Sun, which offers the best sensitivity for spin dependent WIMP-nucleon scattering and is free of alternative astrophysical interpretations. In this talk I will review the status and prospects of the main present and future neutrino telescopes: ANTARES, IceCube and KM3NeT.
\end{abstract}


\bodymatter

\section{Introduction}\label{intro}

A long list of experimental results points to the conclusion that about 85\% of the matter of the Universe is what it is called dark matter, that is, something beyond our known list of particles that does not interact with light, among other known properties. Despite how old the problem is (the first hints were found in 1930s) and the multitude of experiments on this topic, we have not detected it yet.

Given the fact that the properties of the dark matter are not known (or only very incompletely) the experimental effort has to be done in several fronts. These efforts are typically classified in direct detection experiments (in which the dark matter particle interacts in the detector), indirect detection experiments (in which the products of dark matter located in astrophysical sources are observed) and accelerator searches (looking for evidence of missing particles which cannot be explained by known particles).
Among indirect searches, neutrino telescopes are one of the most interesting tools, as explained in the following sections. As an example, a potential signal of high energy neutrinos from the Sun would have dark matter as the most natural explanation, since other astrophysical scenarios would be more exotic. In any case, and as pointed out before, it is important to make sure that all the strategies are used in order to reduce the chances of missing the right candidate.

\section{Experiments}
Neutrino telescopes have become very powerful tools for astroparticle and particle physics. The recent detection of a high energy cosmic neutrino flux by the IceCube collaboration shows that the technique works and has opened a new era in how we observe the Universe.
The detection principle of these instruments is to detect, by means of a three-dimensional array of photomultipliers, the Cherenkov light induced by relativistic leptons produced in charged current interactions of high energy neutrinos. The detection of neutral current interactions is also possible because of the hadronic shower which is also produced. A particularly interesting case is when the flavor of the neutrino is muonic, since the long track of the muon allows for a very good angular resolution. The media where these photomultipliers are installed has to be transparent and cheap (i.e. natural) so the oceans/lakes or the Antarctic ice are the available options. At lower energies, other options like the Super-Kamiokande detector are also competitive (see~\cite{bib_sk} for recent results on the Sun, for instance).

\begin{alphlist}[(d)]
\item ANTARES~\cite{bib_antares} is located in the Mediterranean Sea, 40 km off the French coast, near Toulon. It consists of 885 photomultipliers distributed in 12 lines, which are installed at a depth of 2475 m and kept vertical by buoys. The detector was completed in 2008 and has demonstrated the feasibility of this technique in the sea. One of the advantages of this detector is that water allows for a better angular resolution than ice, since the scattering length is longer. Moreover, being located in the Northern Hemisphere, it can observe more easily the Galactic Center. For the case of the Sun, it also helps to be at intermediate latitudes instead of the South Pole, since the Sun is on average farther away from the horizon, where the atmospheric background is larger.

\item IceCube~\cite{bib_ic} is the first neutrino telescope of a size of about cubic kilometer, which is the “natural” scale for these detectors for the detection of astrophysical sources, as it has been demonstrated by the first detection of cosmic neutrino fluxes. It is an expansion of the AMANDA project. IceCube is made of about 5000 optical modules distributed along 86 lines, including a set of six extra lines that were installed in the inner part of the detector (DeepCore) ot lower the energy threshold. Moreover, given the large size of the detector, the external part can be used as a veto, opening the possibility of looking at the Sun during summer or at the Galactic Center. The installation of the last lines was done in 2010, although physics analyses started with data of configurations with fewer lines during the construction process. 

\item KM3NeT~\cite{bib_jong} is the following step in the field of neutrino astronomy. It will conjugate the advantages of ANTARES (better angular resolution, better location for studying the Galaxy) with the scale of IceCube (actually, about a factor three bigger in its final configuration). Moreover, it will also have a more advanced technology, with the innovative idea of the multi-PMT optical module, i.e. having 31 small photocathode photomultipliers instead of a single large PMT in each optical module. This allows for a better background rejection and more effective area at a reduced cost.

\end{alphlist}

\section{Sources}

If dark matter is made of WIMPs (Weakly Interacting Massive Particles) which can be, for instance, neutralinos (in SuperSymmetry models) or Kaluza-Klein particles (in Universal Extra Dimension models), it would accumulate in astrophysical objects and produce, typically as secondary products, high energy neutrinos.

\begin{alphlist}[(d)]
\item Sun \\
The Sun is probably the best target to look for dark matter with neutrino telescopes. WIMPs would scatter with the nucleons of the Sun (mostly protons), lose energy and become gravitationally trapped. The self-annihilation of these WIMPs would produce neutrinos with energies (tens or hundreds of GeVs) well above the neutrinos produced by nuclear reactions ($\sim$MeV). Therefore, the Sun can be a very clean source, in the sense that no other astrophysical explanations would compete with dark matter. This is in contrast with other indirect searches like those using gammas or cosmic rays (or neutrinos from the Galactic Center).
Limits from several experiments are shown in Figure~\ref{fig_sun}. It can be seen that neutrino telescopes offer the best results for the spin dependent cross section of WIMP-proton scattering. Note also that even if ANTARES is much smaller than IceCube, their results~\cite{bib_zornoza, bib_aartsen} are competitive because of its better angular resolution.

\begin{figure}[h]
\includegraphics[width=\textwidth]{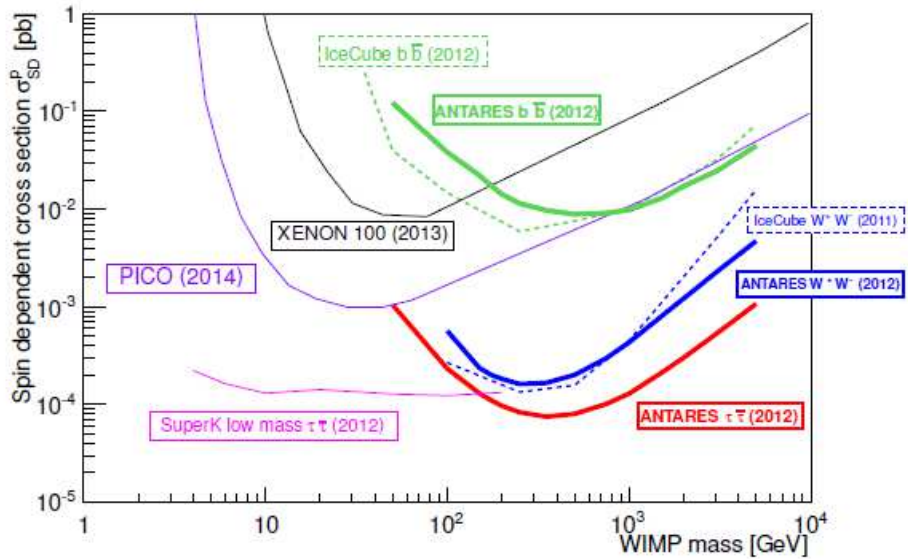}
\caption{Limit on the WIMP-p spin dependent scattering cross section for ANTARES (thick solid): tau-tau (red), W+W- (blue), bb (green), IceCube-79 (dashed), SuperKamiokande (colored dash-dotted), SIMPLE (black short dash-dotted), COUPP (black long dash-dotted) and XENON-100 (black long dashed). Results are compared with a scan in MSSM-7.\cite{bib_icrc}}
\label{fig_sun}
\end{figure}

\item Galactic Centre  \\

The Galactic Center is also an interesting source for neutrino telescopes when looking for dark matter. It obviously has the advantage with respect the Sun that the involved mass is much larger, although located at a larger distance. On the other hand, the high energy component, which is absorbed in the case of the Sun, can arrive better to us from the Galactic Center, being much less dense.
 In this case there is no WIMP-nucleon scattering, so there is only the effect of gravitational accumulation. This is why the limits in this case are set on the WIMP annihilation cross section (averaged with the velocity distribution) as shown in Figure~\ref{fig_gc}. The limits from IceCube~\cite{bib_ackermann} are better than those of ANTARES~\cite{bib_adrian} at low masses, while at high masses is the other way around, since the veto used is not so useful and ANTARES benefits from its better location for observing the Galactic Center.

\begin{figure}[h]
\includegraphics[width=\textwidth]{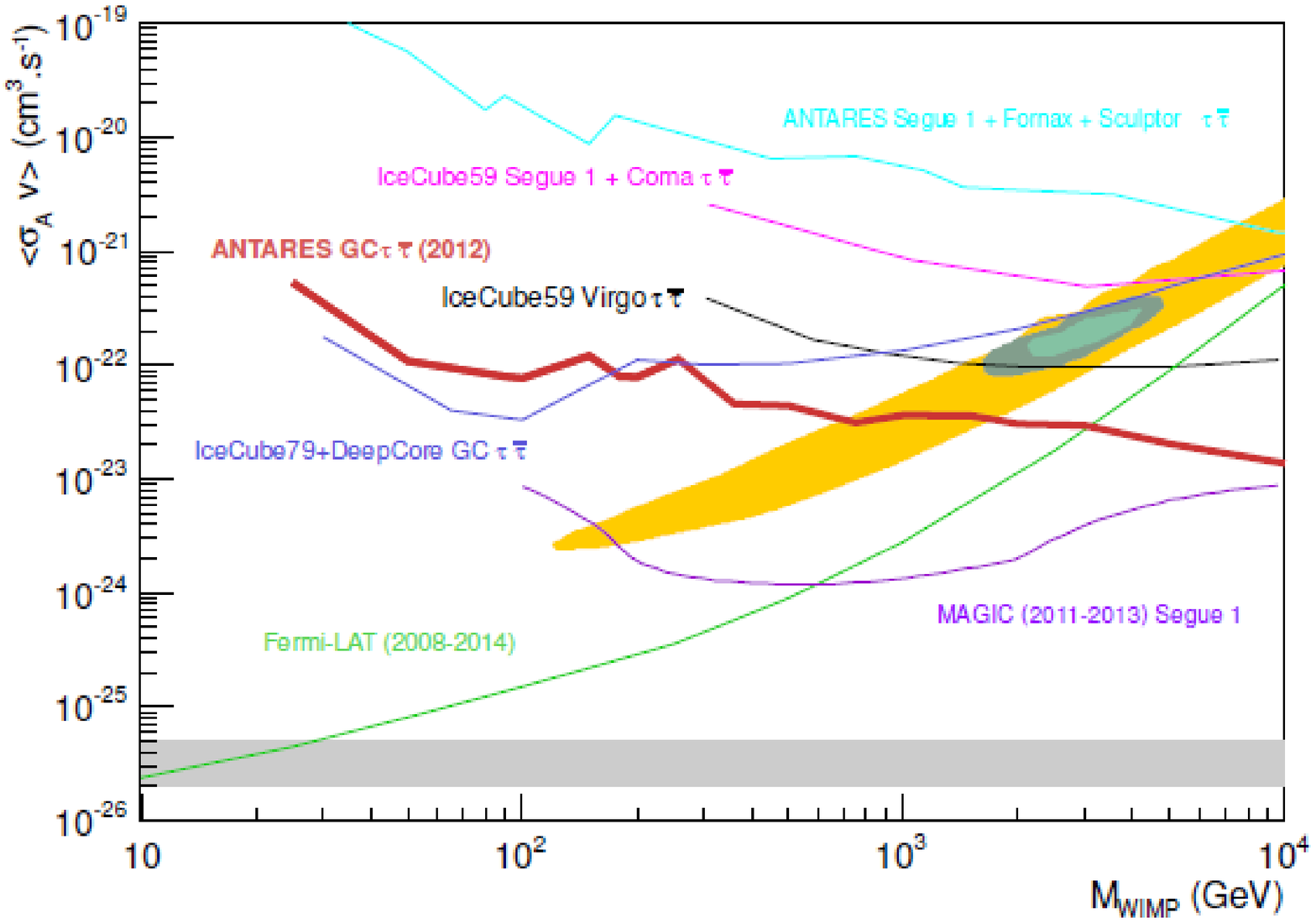}
\caption{Limits on the annihilation cross section averaged on the velocity for different indirect searches\cite{bib_adrian}. The colored areas indicate the dark matter interpretation of the electron/positron excesses of PAMELA, Fermi-LAT and H.E.S.S.\cite{bib_blobs}. The grey band indicates the natural scale for which a WIMP is a thermal relic of the early Universe}
\label{fig_gc}
\end{figure}

\item Other sources \\

Other sources interesting to look at for dark matter are the Earth, the Galactic Halo, dwarf galaxies and galaxy clusters. The case of the Earth is similar to the Sun, with some subtle differences. The interaction is mainly with even nuclei (like iron and nickel) so the limits are set on the spin independent cross section of WIMP-nucleon scattering. Moreover, equilibrium between capture and annihilation cannot be assumed, due to the smaller escape velocity at Earth.
Sources like the Galactic Halo, dwarf galaxies and galaxy clusters are more similar to the case of the Galactic Center (i.e. no WIMP nucleon scattering, so limits on the annihilation cross section are set). The results of some of these searches are also shown in Figure~\ref{fig_gc}.

\end{alphlist}

\section{Conclusions}
Neutrino telescopes have a very wide scientific scope, being the search for dark matter one of the most attractive goals. They have specific advantages with respect to other searches, both when compared with other indirect searches or direct searches, which can be critical when trying to unveil the nature of dark matter, many of the properties of which are still unknown. As an example of the usefulness of these detectors in this context, we can mention the fact that they offer the best limits for spin dependent cross section for WIMP-nucleon scattering. In the case of the Sun, a future detection would be practically background free, which is very important to claim a discovery in this field.

\end{document}